\documentclass[11pt]{article}
\usepackage[dvips,ps2pdf, colorlinks,linkcolor=blue,anchorcolor=blue,citecolor=blue ]{hyperref}
\usepackage{graphicx}
\usepackage{color}
\usepackage{amsmath}
\usepackage{amssymb}
\usepackage{draftcopy}
\draftcopyName{}{50} \draftcopySetGrey{0.8}
\draftcopyPageTransform{0 rotate} \draftcopyPageX{80}
\draftcopyPageY{50}
\usepackage{mathrsfs}
\usepackage{textcomp}
\usepackage{braket}
\usepackage{pifont}
\usepackage{multicol}
\usepackage{bbding}
\usepackage{fancybox}
\usepackage{array}
\usepackage{longtable}
\usepackage{latexsym}
\usepackage{varioref}
\usepackage{xspace}
\usepackage{makeidx}
\usepackage{verbatim}
\usepackage{graphicx}
\usepackage{tabularx}
\usepackage{fancyhdr}
\usepackage{titletoc}

\newtheorem{theorem}{\indent T{\scriptsize HEOREM}}
\newtheorem{lemma}{\indent L{\scriptsize EMMA}}
\newtheorem{definition}{\indent D{\scriptsize EFINITION}}

\newcommand{\xLongLeftRightArrow}[2][]{%
  \ext@arrow 0055{\LongLeftRightArrowfill@}{#1}{#2}}
\def\LongLeftRightArrowfill@{%
 \arrowfill@\Leftarrow\Relbar\Rightarrow}

\title{How to retrieve priced data}
{\author{Li Xi}}

\date{}
\begin{document}
\maketitle
\begin{abstract}
Databases are an indispensable resource for retrieving up-to-date
information. However, curious database operators may be able to
find out the users' interests when the users buy something from
the database. For these cases, if the digital goods have the
identical prices, then a $k$-out-of-$n$ oblivious transfer
protocol could help the users to hide their choices, but when the
goods have different prices, this would not work. In this paper,
we propose a scheme to help users to keep their choices secret
when buying priced digital goods from databases.
\end{abstract}

\begin{center}
\item\section{Introduction}
\end{center}
I am quite sure that all readers are familiar with digital
libraries, such as the digital libraries of ACM, IEEE and SIAM.
These libraries provide researchers with a comprehensive resource
of published papers, and users can easily retrieve their desired
papers by visiting these libraries. Recall how we retrieve data
from digital libraries: we log in the system, select our desired
ones and download them. If one does not own a membership of a
database, he would have to pay for the papers he reads, perhaps
according to the length of the publication. This process is
convenient, but we undertake the risk of revealing our private
research interests to the database operators.

If every paper has the same price, this problem can be resolved
perfectly: suppose that there are $n$ publications in the library
and we are interested in $k$ of them, then by $k$-out-of-$n$
oblivious transfer, we pay for some $k$ publications while
revealing nothing about our choices. In other word, the operator
learns nothing but $k$, and could get the payment by adding up the
prices for the $k$ sold items.

However, it is naive to assume that all publications have the same
price. Nowadays, most papers are priced according to their
lengths, perhaps one dollar per page. A more scientific way
(although not perfect) is to price the data according to the
number of bits it contains. More formally, here we may view the
database as a binary string $x=x_1x_2\cdots x_n$ of length $n$,
and every bit has the same weight. Then the we could still use
oblivious transfer to buy our desired bits from the library,
leaking nothing but the number of bits we pay for.

This scheme is not efficient(since the number of bits can be very
large), and a more serious problem is that we should not assume
every bit to have the same value. Instead of finding a method to
assign prices for different goods, we would rather let the
database operator to assign the value herself himself: a two-page
communication may cost you 100 dollars, while a 200-page review
may cost only 1 dollar. After all, we let the operator to assign
the prices himself.

Now we consider the general problem: the database has $n$ items,
namely $m_1, m_2, \cdots, m_n$. Each item $m_i(1\leq i\leq n)$ has
its own weight $p_i$. Let $\sigma_1, \sigma_2, \cdots, \sigma_k$,
a subset of $\{1, 2, \cdots, n\}$, be the choices of a user, the
goal is to leak $m_{\sigma_j}(1\leq j\leq k)$ to the user while
revealing nothing but $\Sigma_jp_{\sigma_j}$ to the database
operator. This is a special case of oblivious transfer, we denote
it weighted oblivious transfer.

\textbf{Organization.} In the rest of this section we discuss in
more detail traditional oblivious transfer and weighted oblivious
transfer. Section 2 presents two protocols for weighted oblivious
transfer. Section 3 concludes the paper.

\begin{center}
\item\subsection{A short review of oblivious transfer}
\end{center}
Oblivious Transfer ($OT$) refers to a kind of two-party protocols
where at the beginning of the protocol one party, the sender, has
an input, and at the end of the protocol the other party, the
receiver, learns some information about this input in a way that
does not allow the sender to figure out what it has learned
$\cite{israel}$. Oblivious transfer is one of the key components
of many cryptographic protocols and a fundamental primitive for
cryptography and secure distributed computation $\cite{uses,
secure computation, distributed commutation}$. The concept of
oblivious transfer was proposed by Rabin $\cite{rabin}$, since
then, many flavors of oblivious transfer were introduced and
analyzed $\cite{rabin, ot 1-2, ot 1-n, concurrent, 2008}$. Now
oblivious transfer
 is one of the most remarkable achievements in foundation of cryptography.
 The main flavors of oblivious transfer are as follows:

\begin{itemize}
    \item Original oblivious transfer $(OT) \cite{rabin}.$
    For $OT$, the sender has only one
secret, $m$, and would like to have the receiver obtain $m$ with
probability $0.5$. On the other hand, the receiver does not want
the sender to know whether it gets $m$ or not.
    \item 1-out-of-2 oblivious transfer$(OT_2^1) \cite{ot 1-2}.$
    For $OT_2^1$, the sender has two secrets, $m_1$ and $m_2$, and would like to give the receiver one of
them at the receiver's choice. Again, the receiver does not want
the sender to know which secret it chooses.
    \item 1-out-of-$n$ oblivious transfer$(OT_n^1) \cite{ot 1-n}.$
    $OT_n^1$ is a natural extension of $OT_2^1$
     to the case of $n$ secrets, in which the sender has $n$ secrets $m_1, m_2, \ldots, m_n$
      and is willing to disclose exactly one of them to the receiver at its choice.
    \item $k$-out-of-$n$ oblivious transfer$(OT_n^k) \cite{k-n}.$
For $OT_n^k$, the receiver can receive only $k$ messages out of
$n$ messages sent by the sender. In general, one thinks that
$OT_n^k$ is extension of $OT_n^1$. It is obvious that a trivial
$OT_n^k$ protocol can be obtained by performing $OT_n^1$ protocol
$k$ times.

\end{itemize}

Essentially, all these flavors are equivalent in the information
theoretic sense $ \cite{on the foundations}$, but their functions
vary, intuitively, we may use the following relation to describe
the relation among all four flavors:
 \begin{equation}\label{}
    OT\subseteq OT_2^1\subseteq OT_n^1\subseteq OT_n^k.
\end{equation}

There are many ways to construct an efficient oblivious transfer
protocol. Classical oblivious transfer protocols are based on
discrete logarithm $\cite{discrete, Naor}$, the hardness of the
decisional Diffie-Hellman problem $\cite{WG Tzeng}$ \emph{etc}.
\begin{center}
\item\subsection{Definition of weighted oblivious transfer}
\end{center}
To define the requirements of a weighted oblivious transfer
protocol, we simply apply the requirements of general $OT_n^1$
protocol (with minor revisions) $\cite{requirement}$ to it: for
convenience, let $m_1, m_2, \cdots, m_n$ to be items and $p_i$ be
the weight of $m_i$.

\begin{definition}
A $(k$-out-of-$n)$ weighted oblivious transfer should meet the
following requirements:

\begin{itemize}
\item \emph{\textsf{Correctness.}} The protocol achieves its goal
if both the receiver and the sender behave properly. That is, if
both the receiver and the sender follow the protocol step by step,
the receiver gets $m_{\sigma_i}$'s after executing the protocol
with the sender, where $\sigma_i$'s are the receiver's choices,
and the sender learns $\sum_{i=1}^{k}p_{\sigma_i}$(i.e., the whole
price of the goods).

\item \emph{\textsf{Receivers' Privacy-indistinguishability.}} The
transcripts corresponding to the receiver's different choices
$\{\sigma_{ai}\}$ and $\{\sigma_{bi}\}$,
$\{\sigma_{ai}\}\neq\{\sigma_{bi}\}$, are computationally
indistinguishable to the sender if the following equation is
satisfied:
\begin{equation}\label{}
    \Sigma p_{\sigma_{ai}}=\Sigma p_{\sigma_{bi}}.
\end{equation}
If the transcripts are identically distributed, the choice of the
receiver is unconditionally secure.

\item \emph{\textsf{Sender's Privacy-compared with ideal model.}}
We say that the sender's privacy is guaranteed if, for every
possible malicious $R$ which interacts with $S$, there is a
simulator $R'$ (a probabilistic polynomial time machine) which
interacts with $T$ such that the output of $R'$ is computationally
indistinguishable from the output of $R$.
\end{itemize}
\end{definition}

\textbf{Remark.} The weighted oblivious transfer also relies on
the intractability of subset sub problem. The protocol implies
that by the total price of the sold items, the sender cannot tell
which items the receiver bought. Although subset sum problem is
know to be \textsf{NP-complete}, sometimes it is still solvable
(consider the case where the prices are $1, 2, 4, 8, \cdots,
2^{n-1}$, then the binary representation of the total price would
betray the receiver's choice). However, for this case, even if we
apply a trusted third party $T$, the problem still exists. This
problem is solvable when the database is stored by more than one
servers (recall PIR), but this is out of the scope of this paper.

\begin{center}
\item\subsection{Comparing to priced oblivious transfer}
\end{center}

Perhaps the idea that of weighted oblivious transfer is similar to
$\cite{price}$. In $\cite{price}$, the notion of ``priced
oblivious transfer'' is proposed. Informally, assume that a buyer
first deposits a pre-payment at the hands of a vendor. The buyer
should then be able to engage in a virtually unlimited number of
interactions with the vendor in order to obtain digital goods
(also referred to as items) at a total cost which does not exceed
its initial deposit amount. After spending all of its initial
credit, the buyer should be unable to obtain any additional items
before depositing an additional pre-payment. For priced oblivious
transfer, unlimited number of interactions and prepayment is
required, while these requirements relax in this paper. However,
for weighted oblivious transfer, the receiver would disclose
\emph{how much, when} to the sender, since we do not assume that
the receiver interacts with the sender many times. Comparing to
priced oblivious transfer, weighted oblivious transfer is used
when the receiver would like to buy the desired items once at the
same time.

\begin{center}
\item\section{Weighted oblivious transfer}
\end{center}

The idea of our first protocol is straightforward. The sender
locks the ever item $m_i$ with $p_i$ different locks. In this way,
only with all $p_i$ locks can the receiver get $m_i$. This implies
that the sender needs to generate $\sum_{i=1}^np_i$ keys, and with
$\sum_{i=1}^kp_{\sigma_i}$ locks and keys, the receiver could
unlock the locks for $m_{\sigma_i}$'s. By using a
$\sum_{i=1}^kp_{\sigma_i}$-out-of-$\sum_{i=1}^np_i$ oblivious
transfer, the sender could leak the corresponding keys to the
receiver without knowing the chosen ones, thus unable to figure
out the receiver's choices. For simplicity, the best(i.e., most
efficient) lock should be symmetric key encryption scheme. \vskip
2mm
\begin{center}
\textbf{P{\scriptsize ROTOCOL} 1}
\end{center}

In this protocol, the sender has $\sum_{i=1}^np_i$ pairs of
(different) keys, denoted by $K_{ij}(i\in\{1, 2,\cdots, n\},
j\in\{1, 2, \cdots, p_i\})$. Intuitively, the sender locks $m_i$
with $K_{ij}$'s.

\emph{Input}: The receiver's input is composed of $k$ numbers
$\sigma_1, \sigma_2, \ldots, \sigma_k$, which is a subset of $\{1,
2, \ldots, n\}$, and the sender's input is composed of $n$ priced
items $m_1, m_2, \ldots, m_n$, the weight (price) of item $m_i$ is
$p_i$.

\emph{Output}: The receiver's outputs are $m_{\sigma_1},
m_{\sigma_2}, \ldots, m_{\sigma_k}$, and the sender's output is
$\Sigma_{i=1}^kp_{\sigma_i}$.

\begin{itemize}
    \item \textbf{Step 1}
    The sender encrypts the items $m_1,
m_2,\cdots, m_n$ with the encryption keys. For $m_i$, this is done
by computing
$E_{K_{i1}}(E_{K_{i2}}(\cdots(E_{K_{ip_i}}(m_i))\cdots)$. That is,
$m_i$ is encrypted by $p_i$ locks: $K_{i1}, \cdots, K_{ip_i}$
respectively.
\end{itemize}
\begin{itemize}
    \item \textbf{Step 2}
    The sender sends all the ciphertexts to the receiver.
\end{itemize}
\begin{itemize}
    \item \textbf{Step 3}
    By $\sum_{i=1}^kp_{\sigma_i}$-out-of-$\sum_{i=1}^np_i$ oblivious
transfer, the sender reveals the keys for all $m_{\sigma_i}$'s,
while learning nothing about the receiver's choices.
\end{itemize}
\begin{itemize}
    \item \textbf{Step 4}
    With the keys, the receiver easily decrypts and learns all
    $m_{\sigma_i}$'s (and nothing else).
\end{itemize}

This protocol is straightforward, and we do not prove that it is
actually a weighted oblivious transfer in a formal way.
Informally, assume the security of
$\sum_{i=1}^kp_{\sigma_i}$-out-of-$\sum_{i=1}^np_i$ oblivious
transfer protocol used in step 3, the sender leaks nothing but
$\sum_{i=1}^kp_{\sigma_i}$ during step 3, the only communication
from the receiver to the sender. Also, the receiver could unlock
no more than $\sum_{i=1}^kp_{\sigma_i}$ locks, thus leans no more
than what ``costs'' $\sum_{i=1}^kp_{\sigma_i}$. Although the
protocol is not efficient enough, it is the cornerstone of the
next protocol.

\textbf{Remark.} When the protocol is applied by databases, the
first two steps are done before transactions. That is, the
database publishes the encrypted items online and everyone could
download them. When interested in some of the items, the user
interacts with the database operator and completes the last two
steps. In this way, they would not need to communicate the whole
encrypted data, which turns out to be huge. Also, the items are
encrypted only once for all users.

\begin{center}
\item\subsection{Making our protocol efficient}
\end{center}

It is not hard to show that protocol 1 needs $O(\sum_{i=1}^np_i)$
encryptions, this number is clearly impractical, at least
sometimes. In this subsection, we propose a very efficient
protocol which only needs $O(n)$ encryptions.

If for any $m_i$, if there is way to divide $m_i$ into pieces such
that it is easily reconstructable from $p_i$ pieces, but even
complete knowledge of $p_i-1$ pieces reveals absolutely no
information about $m_i$, then we can propose a new protocol. This
is really easy: let
\begin{equation}
m_i=\bigoplus_{j=1}^{p_i}m_{ij}=m_{i1}\oplus
m_{i2}\oplus\cdots\oplus m_{ip_i}
\end{equation}
Then $m_i$ could only be recovered with all $p_i$ $m_{ij}$'s.
However, this division scheme is not very efficient since $m_i$
can be very long. Also, for this case, instead of downloading the
encrypted data from the website, the receiver has to learn all he
needs from the sender. So we slightly revise our idea and comes up
with protocol 2: \vskip 2mm
\begin{center}
\textbf{P{\scriptsize ROTOCOL} 2}
\end{center}
In this protocol, the sender has $n$ pairs of different keys,
denoted by $K_{i}(i\in\{1, 2,\cdots, n\})$. Intuitively, she
intends to encrypt $m_i$ with $K_i$.

\emph{Input}: The receiver's input is composed of $k$ numbers
$\sigma_1, \sigma_2, \ldots, \sigma_k\in \{1, 2, \ldots, n\}$, and
the sender's input is composed of $n$ items $m_1, m_2, \ldots,
m_n$, the weight of item $m_i$ is $p_i$.

\emph{Output}: The receiver's outputs are $m_{\sigma_1},
m_{\sigma_2}, \ldots, m_{\sigma_k}$.

\begin{itemize}
    \item \textbf{Step 1}
    The sender encrypts the items $m_1,
m_2,\cdots, m_n$ with the encryption keys and obtains
$E_{K_1}(m_1), E_{K_2}(m_2),\cdots, E_{K_n}(m_n)$.
\end{itemize}
\begin{itemize}
    \item \textbf{Step 2}
    The sender sends all the ciphertexts to the receiver.
\end{itemize}
\begin{itemize}
    \item \textbf{Step 3}
    For every $i\in\{1, 2, \cdots, n\}$, the sender divides $K_i$ into $p_i$ parts.
    This is done by finding $K_{i1}, K_{i2}, \cdots, K_{ip_i}$
    such that $\bigoplus_{j=1}^{j=p_i}K_{ij}=K_i$.
\end{itemize}
\begin{itemize}
    \item \textbf{Step 4}
    Using oblivious transfer, the sender leaks
    $K_{\sigma_1},\cdots,K_{\sigma_k}$ to the receiver while
    learning nothing about $\sigma_i$'s. This is done by revealing
    $\sum_{i=1}^{k}p_{\sigma_i}$ parts (i.e., the parts of key $K_{\sigma_i}$'s) out of all
    $\sum_{i=1}^{n}p_i$ parts(i.e., for every $\sigma_i$, reveal $K_{\sigma_i1}, \cdots, K_{\sigma_ip_{\sigma_i}}$).
\end{itemize}
\begin{itemize}
    \item \textbf{Step 5}
    The receiver recovers the keys for all $m_{\sigma_i}$'s by
    exclusive-oring $m_{\sigma_ij}$ and decrypts them.
\end{itemize}

Similarly, when the protocol is applied by databases, the first
three steps are done before transactions. That is, the database
publishes the encrypted items online and everyone could download
them. When interested in some of the items, the user interacts
with the database operator and completes the last two steps.

Now we show that protocol 2 is indeed a weighted oblivious
transfer: If both parties behave properly, then the receiver would
learn all parts of $K_{\sigma_i}, i\in\{1, \cdots, k\}$, and by
XOR operations he learns $K_{\sigma_i}$, thus able to learn
$m_{\sigma_i}$.

The scheme takes only three rounds. This is almost optimal since
at least the receiver has to choose $\{\sigma_1, \cdots,
\sigma_k\}$'s and let the sender know and the sender has to
respond to the receiver's request.

For computation, the receiver needs $k<n$ decryptions and
$\sum_{i=1}^kp_{\sigma_i}$ XOR operations. The sender needs $n$
encryptions, $n$ XOR operations and choosing $\sum_{i=1}^n(p_i-1)$
random numbers.

\begin{lemma}
For protocol 2, the receiver's choice is unconditional secure,
assuming that the oblivious transfer used in step 4 is secure.
\end{lemma}

\textbf{Proof.} For any choices $\{\sigma_{ai}\}$, if $\sum
p_{\sigma_{ai}}=\sum p_{\sigma_i}$ holds, in step 4 the receiver
and the sender still perform $\sum p_{\sigma_i}$-out-of-$\sum p_i$
oblivious transfer, then the security of oblivious transfer used
in step 4 shows that the sender cannot learn anything-it cannot
tell $\{\sigma_{ai}\}$ from $\{\sigma_i\}$. Since the receiver
sends nothing else, the sender cannot tell what the receiver's
choice is.$\Box$

\begin{lemma}
For protocol 2, if the receiver is semi-honest, it gets no
information about $m_i, i\notin \{\sigma_1, \cdots, \sigma_k\}$,
assuming the security of the encryption scheme and oblivious
transfer.
\end{lemma}

\textbf{Proof.} If the receiver is semi-honest, due to the
security of oblivious transfer, it learns nothing about $K_i,
i\notin \{\sigma_1, \cdots, \sigma_k\}$. The security of the
encryption scheme promises that $E_{K_i}(m_i), i\notin \{\sigma_1,
\cdots, \sigma_k\}$ is computational indistinguishable from
$E_{K_i}(r), i\notin \{\sigma_1, \cdots, \sigma_k\}$, where $r$ is
a randomly chosen sequence. $\Box$

\begin{lemma}
Protocol 2 meets the requirement of sender's privacy assuming the
security of oblivious transfer used in step 4.
\end{lemma}

\textbf{Proof.} For each malicious receiver $R$ in the real run,
we construct a simulator $R'$ in the Ideal Model such that the
outputs of $R$ and $R'$ are computationally indistinguishable.

As the oblivious transfer used in step 4 is secure, there exists a
simulator $R''$ in the Ideal Model such that the outputs of $R$
and $R''$ are computationally indistinguishable. Now let $R'$ acts
the same as $R''$, then $R$ and $R'$ are computationally
indistinguishable. Since there is no other iteration with $T$, we
prove the theorem.$\Box$

With these preparations, we come up with:

\begin{theorem}
Protocol 2 is indeed weighted oblivious transfer.
\end{theorem}

Since the fact that $OT_n^k$ can be achieved from weighted
oblivious transfer is trivial, we see the equivalence between the
two flavors. Moreover, all (existing) flavors of oblivious
transfer are equivalent in the information theoretic sense.

\begin{center}
\item\subsection{Reducing the computation complexity}
\end{center}

The computational complexity of the protocol is much more
expensive than a traditional $k$-out-of-$n$ scheme where every
item has the same weight. This deficiency may, to some extent,
affect the application of the protocol, but there are some ways to
reduce the computational complexity.

If $p_1,\cdots,p_n$ share a greatest common divisor $q>1$ , then
by dividing $q$ and paying $q$ times the money for each
decryption, the weight of the $i$-th element becomes $p_i/q$,
where $p_i$ is the original weight, and the times of encryptions
and decryptions can be decreased to $1/q$ (of the original one).
In the cases that the greatest common divisor of $p_1,\cdots,p_n$
is 1($q=1$), we introduce two additional methods:

\hskip 1 cm

\textbf{Method 1}\ \ Arrange the items into some certain
categories, and the items of each category share the same weight.
This is practical in everyday life. Assume that there are three
categories, with weight one, two and three, then the computation
is around $3n$.

\hskip 1 cm

\textbf{Method 2}\ \ Generally speaking, methods one could save
most computation. In the cases where it is difficult to arrange
the items into certain categories, we have following method:

First consider an example. Assume that there are four items with
weights 105, 190, 307, 689. Then the greatest common divisor of
the weights is $q=1$, and we could do nothing with them. However,
if the weights change a little into 100, 200, 300, 700, then
$p=100$, and the complexity can be greatly reduced by changing the
weights into 1, 2, 3, 7.

More generally, suppose that the weight of each item is $p_i(i=1,
2, \ldots, n)$. The two parties could decide a ``greatest common
divisor" $q$, and calculate the new weights-the closest integer of
$p_i/q$. In this way, the complexity could also be greatly
reduced.

\begin{center}
\item\section{Concluding remarks}
\end{center}

This paper discusses weighted oblivious transfer, which can be
used for selling priced digital goods. Two implementations of it
was proposed and analyzed. The protocol is especially useful when
the prices of the items are not very large, or the prices of
digital goods fall in very limited categories. In this way, the
computation can be done most efficiently.

However, weighted oblivious transfer also suffers from
shortcomings. We assume that subset problem is hard to compute,
but sometimes it is possible (recall the example that the prices
are $1, 2, \cdots, 2^{n-1}$). And this shortcoming is unsolvable
even when we apply the trusted third party. In addition, sometimes
the whole price of the digital goods itself can leak part (not
all) of the choices, which would also not be secure. Similarly,
the problem exists even a trusted third party is employed. This
additional asks the sender to be careful when assigning the prices
for the goods.

I think that the above problems are unsolvable in current
settings. However, the idea can be used for SPIR, when there are
more than one servers. Further we may consider the implementation
of adaptive queries in the future.

\center


\begin{thebibliography}{99}
\bibitem{israel}M. Naor. Computationally Secure Oblivious
Transfer. \emph{J. Cryptol.}, 18(1), 1-35.

\bibitem{uses}Y. Ishai, M. Prabhakaran, A. Sahai. Founding Cryptography on Oblivious
Transfer-Efficiently. \emph{Advances in Cryptology-CRYPTO 2008,
LNCS 5157}, 572-591, 2008.

\bibitem{secure computation}O. Goldreich, R. Vainish. How to Solve Any Protocol
Problem: An Efficient Improvement. \emph{Advances in
Cryptology-CRYPTO 1987, LNCS 293}, 73-86, 1988.

\bibitem{distributed commutation}J. Kilian. Founding Cryptography on Oblivious
Transfer. \emph{Proceedings of the 20th ACM Symposium on Theory of
Computing}, 20-31, 1988.

\bibitem{rabin}M. Rabin. How to Exchange Secrets by Oblivious Transfer. Technical Report TR-
81, Aiken Computation Laboratory, Harvard University, 1981.

\bibitem{ot 1-2}S. Even, O. Goldreich, A. Lempel. A Randomized Protocol
for Signing Contracts. \emph{Commun. ACM }, 28(6), 637-647.

\bibitem{ot 1-n}G. Brassard, C. Crepeau, J.M. Robert. All-or-Nothing
Disclosure of Secrets. \emph{Advances in Cryptology-CRYPTO 1986,
LNCS 263}, 234-238, 1987.

\bibitem{concurrent}J. Garay, P. MacKenzie. Concurrent Oblivious
Transfer. \emph{Proceedings of the 41th Annual Symposium on
Foudations of Computer Science}, 314-324, 2000.

\bibitem{2008}I. Haitner. Semi-honest to Malicious Oblivious Transfer-The Black-Box
Way. \emph{Proceedings of the Fifth Theory of Cryptography
Conference, LNCS 4948}, 412-426, 2008.

\bibitem{k-n}J. Zhang, Y. Wang. Two provably secure
$k$-out-of-$n$ oblivious transfer schemes. \emph{Appl. Math.
Comput.}, 169(2), 1211-1220, 2005.

\bibitem{on the foundations}C. Cachin. On the Foundations of Oblivious Transfer.
\emph{Advances in Cryptology-EUROCRYPT 1998, LNCS 1403}, 361-374,
1998.

\bibitem{discrete}K. Bruce, L. Cardelli, B. Pierce. Comparing object
encodings. \emph{Inform. Comput.}, 155(1-2), 108-133.

\bibitem{Naor} M. Naor, B. Pinkas. Efficient oblivious transfer
protocols. \emph{Proceedings of the twelfth annual ACM-SIAM
symposium on discrete algorithms}, 448-457, 2001.

\bibitem{WG Tzeng}WG Tzeng. Efficient 1-Out-of-n Oblivious Transfer Schemes with Universally Usable
Parameters. \emph{IEEE Trans. Comput.}, 53(2), 232-240.

\bibitem{requirement} M. Naor, B. Pinkas. Oblivious Transfer and Polynomial
Evaluation. \emph{Proceedings of the 31st ACM Symposium on Theory
of Computing}, 145-254, 1999.

\bibitem{price}B. Aiello, Y. Ishai and O. Reingold. Priced Oblivious Transfer: How to Sell Digital Goods.
\emph{Advances in cryptology-EUROCRYPT 2001, LNCS 2045}, 119-135,
2001.

\end{thebibliography}
\end{document}